\documentclass[a4paper,11pt]{article}
\usepackage{pos}
\usepackage{siunitx}[=v2]
\usepackage[version=4]{mhchem}
\usepackage{tikz}
\usepackage{xcolor}
\definecolor{EarlsGreen}{HTML}{bcbd22}

\usepackage[english]{babel}

\usepackage[subrefformat=parens]{subcaption}
\usepackage{cleveref}
\usepackage{commath}
\usepackage{braket}

\usepackage[
  backend=biber,
  style=phys,
  biblabel=brackets,
  articletitle=false,
  eprint=true,
  url=true,
  giveninits=true,
  maxnames=3
]{biblatex}
\title{Air shower genealogy for muon production}

\hypersetup{
  pdftitle=Air shower genealogy for muon production,
  pdfauthor={Maximilian Reininghaus, Ralf Ulrich, Tanguy Pierog},
  pdflang=en,
  pdfsubject=ICRC 2021
}

\author*[a,b]{Maximilian Reininghaus}
\author[a]{Ralf Ulrich}
\author[a]{Tanguy Pierog}

\affiliation[a]{Karlsruher Institut für Technologie (KIT), Karlsruhe, Germany}
\affiliation[b]{Instituto de Tecnologías en Detección y Astropartículas (ITeDA), Buenos Aires, Argentina}

\emailAdd{reininghaus@kit.edu}

\abstract{
    Measurements of the muon content of extensive air showers at the highest energies show discrepancies compared to simulations as large
    as the differences between proton and iron. This so-called muon puzzle is commonly attributed to a lack of understanding of the hadronic
    interactions in the shower development. Furthermore, measurements of the fluctuations of muon numbers suggest that the discrepancy is
    likely a cumulative effect of interactions of all energies in the cascade. A feature of the air shower simulation code
    CORSIKA~8 allows us to access all previous generations of final-state muons up to the first interaction. With this technique, we study
    the influence of interactions happening at any intermediate stage in the cascade on muons depending on their lateral distance
    in a quantitative way and compare our results with predictions of the Heitler--Matthews model.
}

\FullConference{37$^{\rm{th}}$ International Cosmic Ray Conference (ICRC 2021)\\
		July 12th -- 23rd, 2021\\
		Online -- Berlin, Germany}

\newcommand{\Ngen}{N_{\mathrm{gen}}}

\begin{document}
\maketitle

\newcommand{\Nmu}{N_{\mu}}

\section{Introduction}
Muons in extensive air showers (EAS) induced by cosmic rays (CR) are for the most part the end product of the hadronic
cascade: long-lived, high energy hadrons interact with nuclei of air molecules (mainly \ce{^{14}N} and \ce{^{16}O}),
producing hadrons of lower energy which eventually decay to muons after several generations of interactions. Therefore,
muon observables provide valuable probes of hadronic interactions up to the highest energies. Regarding the total number
of muons $\Nmu$, a discrepancy has been found between EAS simulations using state-of-the-art hadronic interaction models
and measurements by several experiments~\cite{Dembinski:2019uta,Gesualdi:2020ttc}. As the meta-analysis ref.~\cite{Dembinski:2019uta}
shows, this muon defecit (referring to an underestimation of $\Nmu$ in simulations) is present at $\sim 10^{17}\,\si{eV}$,
which corresponds to LHC energy in the center-of-mass frame, and increases with primary energy. It poses a major obstacle to
inferring the mass composition of ultra-high energy cosmic rays (UHECR) from $\Nmu$. On the other hand, a recent measurement of the fluctuations
of $\Nmu$ at the Pierre Auger Observatory shows good agreement between simulations and data~\cite{Aab:2021zfr}. Considering
that this observable is dominated by the first few interactions in the shower, one can reason that the cause of the muon deficit is
likely the effect of accumulating small deviations over several generations covering many orders of magnitude in energy.
To determine which features of hadronic interaction models are good candidates for tweaking to enhance the muon production, it is
beneficial to quantify the relevance of different phase-space regions of hadronic interactions: projectile species, $\sqrt s$,
and kinematic distributions of their produced secondaries.

A similar study has been conducted by \textcite{Hillas:1997tf} with the
computational resources and tools (the EAS simulation code MOCCA) available at that time. A crucial ingredient of this study was the
ability to record and inspect the lineage of particles reaching ground, i.e.\ the mother, grandmother, etc. particles up to the primary.
In contrast to MOCCA and somewhat surprisingly, none of the more recent EAS simulation codes AIRES~\cite{Sciutto:1999jh},
CONEX~\cite{Bergmann:2006yz} and CORSIKA~\cite{Heck:1998vt}, which remain the most widely used ones up until today, provide this feature
to the required extent. Nevertheless, in AIRES and CORSIKA~\cite{Heck:2009zza} mother and grandmother particles are accessible, which
allows to relate the last hadronic interaction in which a secondary meson is produced that subsequently
decays into the muon with that muon~\cite{Meurer:2005dt}.

For the currently being developed EAS simulation code CORSIKA~8~\cite{Engel:2018akg} we designed and implemented an algorithm
capable of retaining the complete lineage of each particle. The information available about any ancestor particle include the type of event in which
the particle was produced (decay or interaction), the state of the projectile (position, 4-momentum) of that event and the state
of all secondaries. It is also foreseen (though not implemented yet) to include dynamical meta-information of events. For example,
for hadronic interactions it may be desirable to save quantities like the number of wounded nucleons, elasticity or whether or not the
interaction was diffractive. A more detailed technical description is given in ref.~\cite{Alves:2021}.

In this work, we make use of this feature in simulations of UHECR-induced EAS ($10^{17}\,\si{eV}$ and above)
with CORSIKA~8 to study hadronic interactions happening throughout the whole shower development with respect to final-state
muons on ground.

\section{Methods}
We simulate EAS with a hybrid approach. We use CORSIKA~8 for a Monte Carlo treatment of
the hadronic and muonic shower components, while electromagnetic (EM) particles, mainly the product
of $\pi^0$ and $\mu^{\pm}$ decays, are fed into CONEX to generate EM longitudinal profiles by numerically
solving the cascade equations. Photo-production of hadrons is not taken into account. We use the hadronic
interaction models SIBYLL~2.3d~\cite{Engel:2019dsg} and QGSJetII-04~\cite{Ostapchenko:2010vb} for energies
above $10^{1.8}\,\si{\GeV} = \SI{63.1}{\GeV}$. Below that energy and down to the cutoff of \SI{631}{\MeV}
kinetic energy we use the Hillas splitting algorithm~(HSA)~\cite{Hillas:1981} as implemented in an extended
version in AIRES~19.04.00~\cite{Sciutto:1999jh} that we link to. The inelastic cross-sections in this energy range are taken from tabulated
values that are part of the UrQMD~\cite{Bleicher:1999xi} distribution in CORSIKA~7. The atmospheric model used
is that of Linsley (see e.g.\ ref.~\cite{CruzMoreno:2013}). The observation level is placed at \SI{1400}{m}~a.s.l.
All data presented are averaged over 400~showers.

\section{Results}
\subsection{Interaction spectrum}

\begin{figure}
\centering
	\input{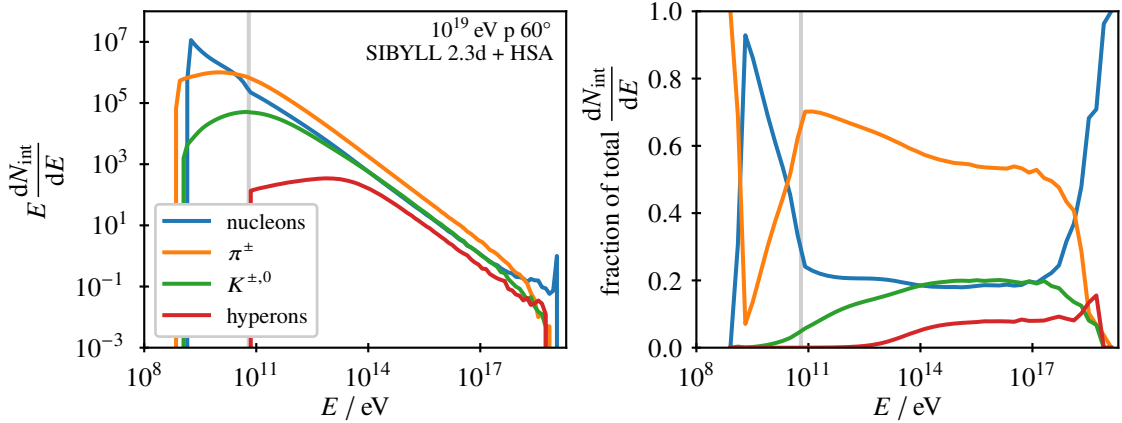}
	\caption{Number of hadronic interactions by energy and species. The grey vertical line indicates the transition
	between low- and high-energy interaction models.}
	\label{fig:dNintdE}
\end{figure}

\Cref{fig:dNintdE} shows the \emph{interaction spectrum}, i.e.\ the number of hadronic interactions
by energy, of a $10^{19}\,\si{\eV}$ proton shower of \ang{60} zenith angle, grouped by several classes of hadrons. This observable
is related to the corresponding energy spectra by
\begin{equation}
\dod{N_{\mathrm{int}}}{E} = \frac{1}{\lambda_{\mathrm{int}}(E)} \int \dif X \dod{N}{E}.
\end{equation}
The interaction spectrum of a given particle species is mainly influenced by the multiplicity
of this species as secondaries in hadronic interactions, as well as its critical energy.

Going from high to low energies, we first observe a peak at the primary energy, which in the limit of infinitesimal
bin widths would be a delta function. For about one decade in energy below the primary energy, most interactions
are those of nucleons, which can be attributed to the leading nucleons of the primary interaction. Below the
crossing point, pion-air interactions are the dominating component, making up \SIrange{50}{70}{\percent} of all
hadronic interactions. Between $10^{14}\,\si{\eV}$ to $10^{17}\,\si{\eV}$ kaons and nucleons contribute equally
to the total interaction spectrum with about \SI{20}{\percent} each. Below $10^{14}\,\si{\eV}$, the kaon contribution
decreases due to the $K^0_S$ starting to predominantly decay instead of interacting. At that point also hyperons (mainly
$\Lambda/\bar\Lambda$, the most long-lived hyperon), which are generally rare and rinteract only to a minor extent at all, fade away almost entirerly.
For a wide range in energy the total spectrum as well as the individual components follow a power-law.
Performing a linear fit of the total $\log(E \mathrm d N_{\mathrm{int}}/\mathrm d E)$ vs.\ $\log E$ in the range
\SI{1}{\TeV} to \SI{0.1}{EeV}, we obtain an exponent of \num{-0.890(2)}. The individual power-laws of unstable hadrons
are broken when the corresponding species reaches its critical energy. Around \SI{100}{\GeV} the most long-lived
and down to this energy most abundant $\pi^\pm$ also start to drop out. The switch from high- to low-energy interaction
model causes a sudden change in the nucleon spectra, which is an artifact of the simplified treatment of the HSA.
These low-energy nucleon interactions play only a minor role regarding muon production, however, as we will show later.

\subsection{Number of generations}

\begin{figure}
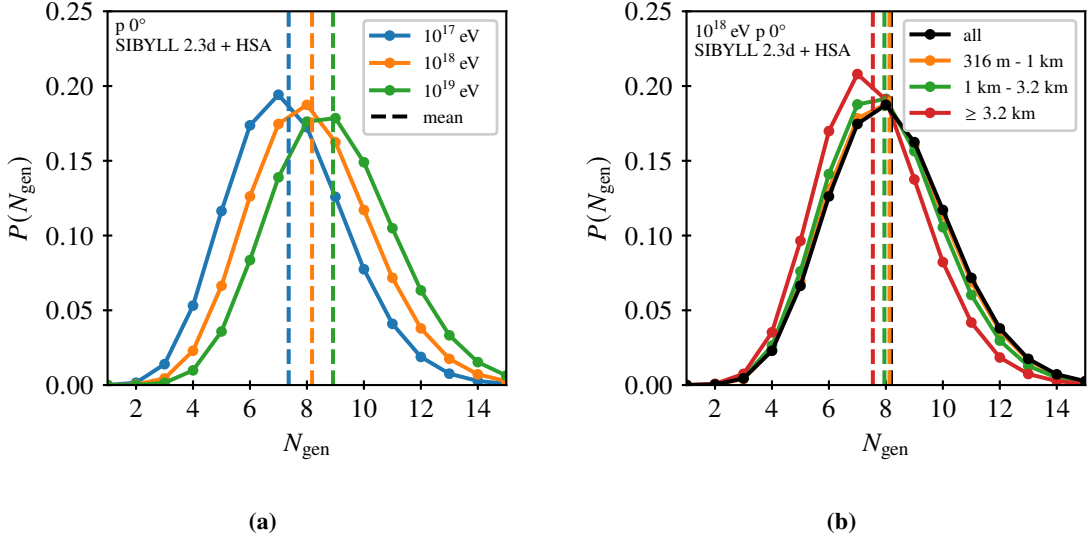

\begin{subfigure}[c]{0.5\textwidth}
	\input{figures/generations-by-E0_p_0deg_sib23d+HSA.pgf}
    \subcaption{}\label{fig:Ngen_E0}
\end{subfigure}
\begin{subfigure}[c]{0.5\textwidth}
	\input{figures/generations-by-radius_p_1e18_0deg_sib23d+HSA.pgf}
    \subcaption{}\label{fig:Ngen_r}
\end{subfigure}
\caption{Number of muon ancestor generations \subref{fig:Ngen_E0} by primary energy \subref{fig:Ngen_r} grouped by radial ranges around shower core.
Vertical dashed lines indicate the mean value of the distribution of the same colour.}
\label{fig:generations}
\end{figure}

In \cref{fig:generations} we study the number of generations $\Ngen$ of ground-reaching muons, which is the total number
of hadronic interactions that connect the primary particle with the muon in the shower. It is an important quantity since
the number of muons grows exponentially with $\Ngen$~\cite{Matthews:2005sd,Cazon:2018gww} and small changes in hadronic interactions, e.g.\ the
energy fraction transferred from the projectile onto further long-lived hadronic secondaries, are correspondingly amplified $\Ngen$ times by the multiplicative
process~\cite{Cazon:2020jla,Albrecht:2021yla}. Making use of the lineage technique, $\Ngen$ is obtained by iterating over the muon ancestors and
counting only interaction events (in contrast to decays). \Cref{fig:Ngen_E0} shows the distributions for different primary energies. The mean value grows
logarithmically with the primary energy as expected from the Heitler--Matthews model~\cite{Matthews:2005sd}. A linear fit of $\braket{\Ngen}$ vs. $\log_{10}(E)$, in which we include
also data of $10^{17.5}\,\si{\eV}$ and $10^{18.5}\,\si{\eV}$ showers not shown in the plot, yields an increase of $\Ngen$ of $s = \num{.785(17)}$ per decade of energy.
In the Heitler--Matthews model, $s$ is related to the hadron multiplicty $m$ via $s = 1 / \log_{10}(m)$, so that we can derive $m = \num{18.8(12)}$.
In \cref{fig:Ngen_r} we consider only muons within certain radial ranges $r$ around the shower core. We observe that muons further away from
the core tend to have slightly fewer generations than those close to the core.

\begin{figure}
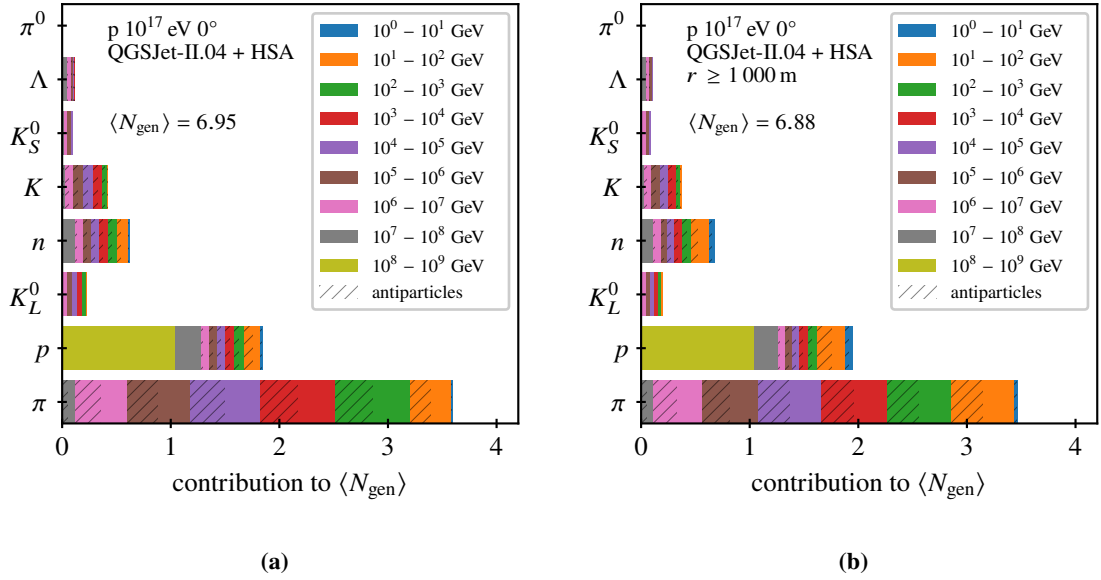

\begin{subfigure}[c]{0.5\textwidth}
    \input{figures/interactions_by_species_qgs2_p-1e17.pgf}
    \subcaption{}\label{fig:contribNgen_all}
\end{subfigure}
\begin{subfigure}[c]{0.5\textwidth}
	\input{figures/interactions_by_species_qgs2_p-1e17_r1km.pgf}
    \subcaption{}\label{fig:contribNgen_1km}
\end{subfigure}
\caption{Distribution of muon ancestor projectile generations by energy and species \subref{fig:contribNgen_all} all muons \subref{fig:contribNgen_1km}
only muons with $r \geq \SI{1000}{\metre}$.}\label{fig:contribNgen}
\end{figure}

It is instructive to quantify to which degree interactions in certain energy ranges and with certain projectiles contribute to the total $\Ngen$.
A priori we can only state the obvious: The first interaction, being the root of the shower, contributes exactly one generation. We build
a histogram binned in projectile energy and species by iterating over the muon lineages and filling the histogram for each interaction according
to its projectile energy and species. Thereby each muon increases the total histogram count by its individual $\Ngen$. Since muons share parts of
their lineage, the corresponding interactions are counted multiple times -- their \emph{muon weight} is given by the number of muons stemming
from that interaction. If we finally divide the bin counts by the total number of muons (possibly after applying a section criterion, e.g.\ on $r$),
we end up with that bin's contribution to $\braket{\Ngen}$. The result of that procedure is shown in \cref{fig:contribNgen}, applied to $10^{17}\,\si{\eV}$ showers.
As expected, the bin containing solely the primary interaction (\tikz{\fill[color=EarlsGreen] (0,0) rectangle (1ex,1ex);}-coloured) has a value of one.
In the energy decade below the primary energy, the main contribution is due to nucleons with about twice as many protons as neutrons, conforming 
with \cref{fig:dNintdE}. In this energy range virtually no contribution of antinucleons is apparent. Charged pions contribute approximately half
of the total $\braket{\Ngen}$. Each $\log E$ range between \SI{10}{PeV} and \SI{100}{GeV} carries comparable weight, slightly decreasing with energy.
No distinction between positively and negatively charged pions can be observed. Below \SI{100}{GeV} the importance of pion interactions decreases again
as more and more pions do not reinteract. Comparing the distributions obtained when selecting only muons with at least \SI{1}{\km} lateral distance (\cref{fig:contribNgen_1km})
with those without any cut (\cref{fig:contribNgen_all}), we find that for great lateral distance the importance of low-energy interactions increases.
This can be understood considering that typically muons with higher energies stay close to the shower core. As the projectile energy of the last
interaction of these muons needs to be higher than the final muon energy, the phase space that can contribute to these muons is necessarily cut off earlier.

\subsection{Pseudorapidity distributions}
\begin{figure}
\centering
\input{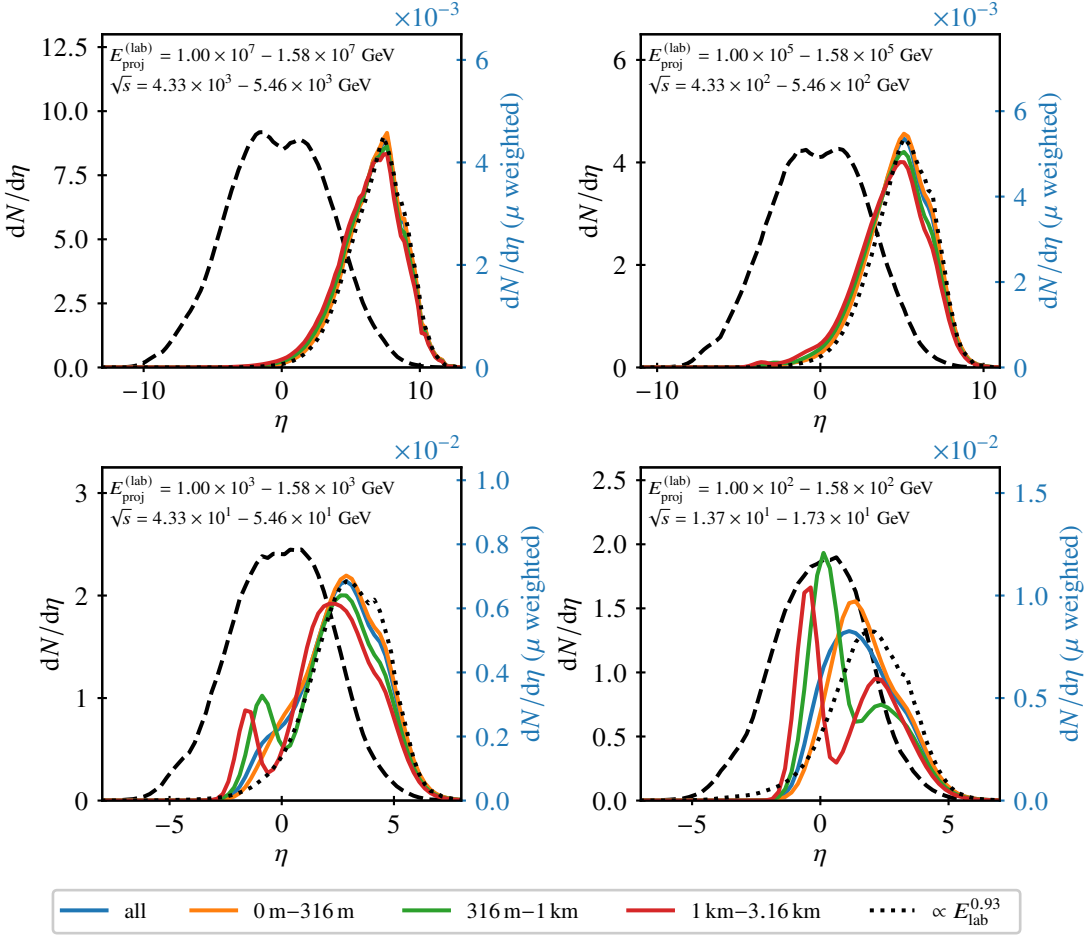}
\caption{Pseudorapidity distributions of $\pi^\pm + \text{Air} \rightarrow \text{charged hadrons}$, $10^{19}\,\si{\eV}$ vertical proton.
The dashed line indicates generator-level distributions while the coloured and dotted lines shows muon-weighted distributions (in arbitrary units).}
\label{fig:eta}
\end{figure}

To quantify the importance of different regions of the phase-space of secondaries in hadronic interactions for muon production,
one may weight specific phase-space element by the number of muons descending from particles produced in it. A simple
prescription based on the Heitler--Matthews model is to weight by $E_{\mathrm{lab}}^\beta$ (with $\beta = \num{.93}$, $E_{\mathrm{lab}}$ is
the energy of the secondary in the lab frame)~\cite{Cazon:2018gww,Albrecht:2021yla}. This, however, does not allow for any
cut to be applied on the muons, e.g.\ on lateral distance. Having the full lineage available in our Monte Carlo simulations,
we follow the approach of \textcite{Hillas:1997tf} and obtain the weight by counting the muons as described in the previous
section. In \cref{fig:eta}, we show pseudorapidity distributions $\mathrm d N/ \mathrm d\eta$ of $\pi^\pm + \text{Air} \rightarrow
\text{charged hadrons}$ (in center-of-mass frame) for four different energies. The pure generator-level distributions (generated with SIBYLL~2.3d in CRMC~\cite{CRMC}
for a fixed projectile energy $E_{\mathrm p}$ and \ce{^{14}N} target) are plotted with dashed lines. The corresponding muon-weighted distributions are obtained from simulations
of vertical $10^{19}\,\si{\eV}$ proton showers in which we consider interactions within a range around $E_{\mathrm p}$. The solid, coloured
lines indicate the weighted distributions (in arbitrary units) after applying a lateral distance cut and normalized by the number of
muons selected.  Additionally, the black solid line shows the $E_{\mathrm{lab}}^\beta$-based weighting for comparison. We find that
at lab energies $\gtrsim 10^{14}\,\si{\eV}$ the weighted distributions almost coincide irrespective of the muon lateral distance. Furthermore,
the $E_{\mathrm{lab}}^\beta$-based weight agrees very well with the distributions (up to an arbitrary scaling factor). These
results quantitatively demonstrate the importance of the forward region of hadronic interactions for muon production.
At lower energies, on the other hand, the muon lateral distance has an impact on the corresponding weight distributions.
Besides the peak in the forward region, a second peak at mid-rapidity around $-2 \lesssim \eta \lesssim 0$ emerges
when only muons with at least a few hundred meters distance are considered, which is not described by the $E_{\mathrm{lab}}^\beta$-based weighting.

\section{Conclusions}
In this work, we have studied the lineage of muons in air shower simulations of UHECR protons, consisting of hadronic interactions.
We have shown that the average number of generations $\Ngen$ grows logarithmically with the primary energy in accordance with
the Heitler--Matthews model. For large lateral distances, the distribution of $\Ngen$ shift towards lower values. The typical muon lineage
contains at energies close to the primary energy mostly interactions of nucleons. The remaining energy range is dominated by pion interactions.
Furthermore, we have applied a "muon-weighting" to pseudorapidty distributions of pion-air interactions, showing the importance
of the forward region in a quantitative manner for high energy interactions. At lower center-of-mass energies, say $\sqrt s \lesssim \SI{100}{\GeV}$,
and especially for muons at large lateral distances also the region at lower values of $\eta$ becomes more and more relevant.

Accelerator measurements conducted in a much more controlled environment that cover the relevant phase-space and emulate the interactions
in EAS as closely as possible are key ingredients to constrain the hadronic interaction models better. Fixed-target experiments such as
NA61/SHINE with $\pi^\pm$ beams are especially relevant to study the last generation of hadronic interactions. Complementary to that are LHC measurements,
ideally of proton-oxygen collisions~\cite{Citron:2018lsq}, which constrain the most energetic interactions in ultra-high energy EAS. Here, data of
forward measurements are highly valued input.

\acknowledgments
The simulations were performed on the bwForCluster BinAC of the University of Tübingen, 
supported by the state of Baden-Württemberg through bwHPC and the DFG through grant
no.\ INST~37/935-1~FUGG.

\printbibliography

\end{document}